\documentclass[doublecol]{epl2} 
\usepackage{graphicx}
\usepackage[fleqn]{amsmath}
\usepackage{verbatim}

\title{Why is the ground state electron configuration for Lithium $1s^22s$~?}
\author{W.S. Stacey\inst{1} \and F. Marsiglio\inst{1,2}}
\shortauthor{W.S. Stacey and F. Marsiglio}
\institute{                    
  \inst{1} Department of Physics, University of Alberta, Edmonton, Alberta, Canada, T6G~2E1\\
  \inst{2} Physics Division, School of Science and Technology, University of Camerino, I-62032 Camerino (MC), Italy
}
\abstract{
The electronic ground state for Lithium is $1s^22s$, and not $1s^22p$. The traditional argument for why this is so
is based on a screening argument that claims that the $2p$ electron is better shielded by the $1s$ electrons, and therefore
higher in energy then the configuration that includes the $2s$ electron. We show that this argument is flawed, and in
fact the actual reason for the ordering is because the electron-electron interaction energy is higher for the 
$2p-1s$ repulsion than it is for the $2s-1s$ repulsion.}

\begin{document}

\maketitle
At the heart of our understanding of atoms, molecules, and solids is an interplay between electron-ion,
electron-electron, and ion-ion interactions. Bound states exist {\em in spite} of repulsive ion-ion interactions
so it is really the first two interactions that collaborate to produce conglomerate forms of matter.\cite{remark0}
If we confine our discussion to atoms, then it is clear from the structure of the periodic table that a significant
(and appropriate) bias has been applied in favour of the electron-ion interactions.\cite{remark_bloch} 
Historically this is no doubt due to two reasons: first, the single electron problem, where electron-electron 
interactions are absent,
has an exact solution (so it is easy to `conceptualize' bound states of many-electron atoms in terms of these
solutions), and second, as the atomic number $Z$ of the nucleus increases, the energy 
associated with the electron-ion interaction (and with the electron kinetic energy) grows as $Z^2$, and is therefore
dominant. Thus, for
the most part, the characteristics of electronic properties of the elements of the periodic table are
described in terms of ``Hydrogenic states,'' i.e. states in which electron-electron interactions are absent. 

Nonetheless electron-electron interactions can have an important effect on the properties of elements,
particularly when they are introduced into a solid state environment.\cite{remark1} 
In this paper we wish to re-examine a
particularly simple case, that of the ground state electron configuration in Lithium. The electrons in Lithium
are said to reside in the $1s^22s$ (i.e. $^2S$) state, which is 
standard short-hand for the antisymmetrized state that
is more fully described below. In the absence of electron-electron interactions, the $1s^22s$ and 
$1s^22p$ (i.e. $^2P$) states for Lithium are degenerate. An accurate computation, utilizing Hylleraas basis 
states,\cite{yan95} makes it clear that when interactions are fully taken into
account, the $^2S$ state indeed has a lower energy, as is clearly observed.

The standard argument for why this is the case proceeds as follows.\cite{griffiths05} The higher angular momentum of the $2p$ electron in the $^2P$ state tends to ``push" the $2p$ electron further out from
the nucleus than is the case with the $2s$ electron in the $^2S$ state. This is based on a semi-classical
explanation that invokes the centrifugal force associated with the higher angular momentum of the
electron. Two opposing effects now arise. Being further from the centre, the $2p$ electron interacts less with the inner $1s$ electrons; this lowers the energy. However, the $2p$ electron is also screened more from the nucleus, and cannot take full advantage of the electron-ion interaction; this therefore raises the energy. 
The effects of screening are argued to outweigh the lessened interaction with the inner electrons, and therefore, the $^2S$ is the preferred state.

While both effects are rooted in the electron-electron interaction, the screening argument puts
emphasis on the energy lowered through the electron-ion interaction, while the opposing argument
clearly emphasizes the direct electron-electron repulsion between the various electrons.\cite{remark2}
In particular, for a small number of particles, and when distances are not large, the distinction between screening and electron-electron interactions is not clear. In this paper, we address this issue and 
reexamine the traditional argument for Lithium's preferred state. We will make the claim that it is flawed;
contrary to the standard argument espoused in the previous paragraph, screening is somewhat irrelevant and it is the 
direct electron-electron repulsion between electrons that plays a central role in deciding the ground state.

The first problem with the standard argument is the claim that ``the $2p$ electron is further out from the nucleus
than the $2s$ electron.'' This is incorrect. It is true that the $2p$ state has a node at the origin, but Kramer's relation\cite{griffiths289} allows one to compute the expectation values
\begin{equation}
<r>_{n\ell} = {a \over 2}\bigl[3n^2 - \ell (\ell + 1) \bigr],
\label{r1}
\end{equation}
and
\begin{equation}
<r^2>_{n\ell} = {n^2 a^2 \over 2}\bigl[5n^2 - 3 \ell (\ell + 1) + 1 \bigr].
\label{r2}
\end{equation}
From either it follows that the expectation value of the radial extent of the wave function is larger for the $2s$
than for the $2p$ state, contrary to the semi-classical explanation. One can argue that expectation values of the
inverse powers of the radial coordinate are more important for the energetics. Then $<1/r> = 1/(n^2a)$ doesn't
discriminate between the two, while $<1/r^2>  = {1 \over n^3a^2}{1 \over \ell + 1/2}$ does indeed indicate that
the $2p$ state is more extended. 

To settle this argument in an unbiased fashion, we follow 
Ref. [\cite{jugdutt12}], and solve for the radial wave function of a particle in an effective potential (Coulomb +
centrifugal term), using the one dimensional radial equation. The method of solution presented in the preceding
reference requires the use of an infinite square well potential, extending from $r=0$ to $r=a_{\rm wall}$, to `embed'
the potential of interest. For sufficiently large $a_{\rm wall}$ the low energy eigenvalues and eigenstates are
insensitive to the existence of this outer wall.\cite{remark3} However, in the present work, in order to see which of the
two states has larger extension, as measured by the energy, we simply have to lower $a_{\rm wall}$ and monitor
which of the two energies, $E_{2s}$ or $E_{2p}$ is affected more. Fig.~1 shows the result of such a calculation,
for both the $2s$ and $2p$, and for the $3s$, $3p$, and $3d$ states. There is no question that in either case 
the energy of the $s$ state is the most affected by the presence of the embedding potential, and hence the 
most extended. This is in agreement with the expectation based on Eq. (\ref{r1}) or (\ref{r2}). In this sense the $2s$ state in
Lithium is more susceptible than the $2p$ state to being screened by the inner $1s$ electrons. Thus, if the naive argument
advanced above were correct, the ground state would be primarily $1s^22p$, not as observed. We now examine the
interactions of the `outer' $n=2$ electron with the $1s$ electrons.

%figure 1
\begin{figure}[h!]
\begin{center}
\includegraphics[height=3.3in,width=3.3in,angle=-90]{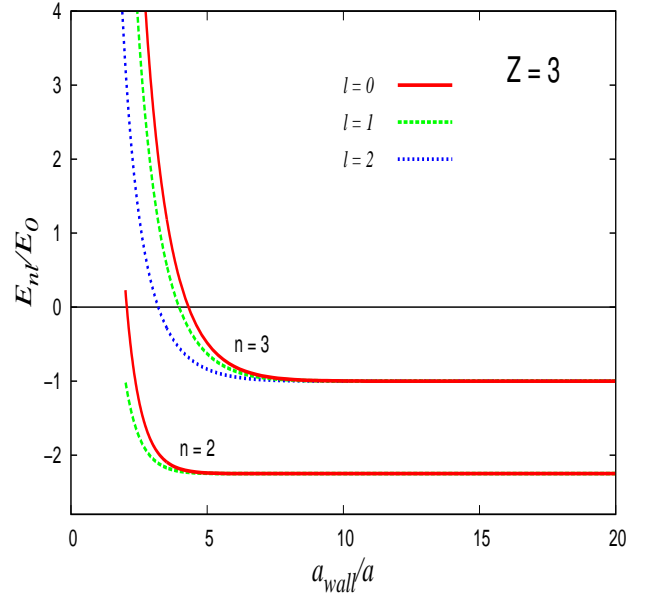} 
\caption{(Colour online) Energy of the $2s$ vs. the $2p$ states (lower two curves) and $3s$ vs. $3p$ vs. $3d$ states
(upper three curves) calculated as a function of the width of an embedding infinite square well potential. For large width, the
states are degenerate as expected for the Coulomb potential, but as the width is lowered, the first state affected in
either case is the $s$-state, indicating that it is the more extended of the degenerate set.
}
\label{fig1}
\end{center}
\end{figure}

We first use a simple variational wave function to calculate the energies of the $1s^22s$ and $1s^22p$ states in Lithium.
Our trial wavefunction has the usual determinant form appropriate to the case where we ignore interactions.
It is
\begin{equation}
 \Psi(\mathbf{r}_1, \mathbf{r}_2, \mathbf{r}_3) =
\frac{1}{\sqrt{6}} \left|
\begin{matrix}
    f(\mathbf{r}_1)\alpha(\mathbf{r}_1) & f(\mathbf{r}_1)\beta(\mathbf{r}_1)  & g(\mathbf{r}_1)\alpha(\mathbf{r}_1) \\
   f(\mathbf{r}_2)\alpha(\mathbf{r}_2)  & f(\mathbf{r}_2)\beta(\mathbf{r}_2) & g(\mathbf{r}_2)\alpha(\mathbf{r}_2) \\
    f(\mathbf{r}_3)\alpha(\mathbf{r}_3) & f(\mathbf{r}_3)\beta(\mathbf{r}_3) &g(\mathbf{r}_3)\alpha(\mathbf{r}_3)
\end{matrix} \right| 
\end{equation}
where
\begin{equation}
f(\mathbf{r}) = \sqrt{\frac{z_1^3}{\pi a^3}} \exp{\left(-\frac{z_1 r}{a}\right)}
\label{ffunc}
\end{equation}
and
\begin{equation}
g(\mathbf{r}) =
\begin{cases}
 \sqrt{\frac{z_2^3}{8 \pi a^3}} \left(1-\frac{z_2 r} {2 a}\right) \exp{\left(-\frac{z_2 r}{2a}\right)}, & {1s^22s} \\
 \sqrt{\frac{z_2^3}{32 \pi a^3}} \frac{z_2 r} { a} \exp{\left(-\frac{z_2 r}{2a}\right)} \cos{(\theta)}, & {1s^22p} 
\end{cases}
\label{gfunc}
\end{equation}
Here $a$ is the Bohr radius, and $\alpha(\mathbf{r})$ and $\beta(\mathbf{r})$ represent the usual spin states. For Lithium, $z_1=z_2=3$; here they are retained as variational parameters. We use this trial wavefunction to place an upper limit 
on the energy of each state. The nonrelativistic Hamiltonian for the Lithium atom is 
\begin{equation}
\begin{split}
H = H_{kinetic} + H_{nuclear} + H_{interaction} \phantom{aaaaaaaaaa} \\ 
= \frac{-\hbar^2}{2m}(\nabla_1^2 + \nabla_2^2 + \nabla_3^2) - \frac{e^2}{4 \pi \epsilon_o}\left(\frac{Z}{r_1} + \frac{Z}{r_2} + \frac{Z}{r_3}\right) \\
 + \frac{e^2}{4 \pi \epsilon_o} \left( \frac{1}{|\mathbf{r}_1 - \mathbf{r}_2|} + \frac{1}{|\mathbf{r}_1 - \mathbf{r}_3|} + \frac{1}{|\mathbf{r}_2 - \mathbf{r}_3|}\right) 
 \end{split}
\label{ham}
\end{equation}
where the terms corresponding to each part of the Hamiltonian should be clear.
Note that $Z$ is the nuclear charge. 
Contributions to the energy are divided into the following categories: $K \equiv <H_{\rm kinetic}>$,  $U_{\rm nuclear} \equiv
<H_{\rm nuclear}>$, and $U_{1s1s}$, $U_{1s2s}$, $U_{1s2p}$, $Ex_{1s2s}$, and $Ex_{1s2p}$ refer to the direct Coulomb
and exchange energies for the orbitals listed.
Then the total energies for each state are
\begin{eqnarray}
E_S &=& {K + U_{\rm nuclear} + U_{1s1s} + 2 U_{1s2s} + Ex_{1s2s}} \label{1s2s} \\
E_P &=& {K + U_{\rm nuclear} + U_{1s1s} + 2 U_{1s2p} + Ex_{1s2p}}.
\label{1s2p}
\end{eqnarray}
Calculations are straightforward; with $E_0 \equiv \hbar^2/(2ma^2)$ ($=1$ Rydberg), the three terms in common for the two states are
\begin{equation}
{K \over E_0} = 
%\int \! \bar \Psi H_{kinetic} \Psi \, \mathrm{dr}_1^3\mathrm{dr}_2^3\mathrm{dr}_3^3 = \frac{\hbar^2}{2 m a^2} 
2z_1^2 + \frac{1}{4}z_2^2, 
\label{kinetic}
\end{equation}
\begin{equation}
%&U_{nuclear} = \int \! \bar \Psi H_{nuclear} \Psi \mathrm{dr}_1^3\mathrm{dr}_2^3\mathrm{dr}_3^3 \\
{U_{\rm nuclear} \over E_0} = -Z \left(4z_1 + \frac{1}{2}z_2\right)
\label{unuclear}
\end{equation}
and
\begin{equation}
{U_{1s1s} \over E_0} = \frac{5}{4}z_1.
\label{u1s1s}
\end{equation}

Then for the two trial states, we have
\begin{eqnarray}
&&{U_{1s2s} \over E_0}= 2 \int  |f({\bf r_1})|^2 \frac{a}{|\mathbf{r}_1 - \mathbf{r}_2|}  |g({\bf r_2})|^2 \, \mathrm{d^3r_1}\mathrm{d^3r_2} \nonumber \\
&&= 2z_1z_2 \frac{(8z_1^4 + 20z_1^3z_2 + 12z_1^2z_2^2 + 10z_1z_2^3 +z_2^4)}{(2z_1+z_2)^5},
\label{u1s2s}
\end{eqnarray}
and
\begin{eqnarray}
{U_{1s2p} \over E_0} &=& 2 \int |f(\mathbf{r}_1)|^2 \frac{a}{|\mathbf{r}_1 - \mathbf{r}_2|}  |g(\mathbf{r}_2)|^2 \mathrm{d^3r_1}\mathrm{d^3r_2} \nonumber  \\
&=& {U_{1s2s} \over E_0} +\frac{16z_1^3z_2^3}{(2z_1+z_2)^5},
\label{u1s2p}
\end{eqnarray}
respectively.
The exchange terms ($Ex_{1s2s}$ and $Ex_{1s2p}$) arise from consideration of the full antisymmetric wavefunction. 
For the two trial states, the exchange terms are
\begin{eqnarray}
&&{Ex_{1s2s} \over E_0} =-8192z_1^4z_2^3\Bigg\{\frac{(z_1-2Z)(z_1-z_2)^2}{4(2z_1+z_2)^8} \nonumber \\
&&+ \frac{(z_1-z_2)(z_1z_2(4z_1-z_2) - Z(4z_1^2-z_2^2))}{8z_1(2z_1+z_2)^8} \nonumber \\
&&+ \frac{(z_1-z_2)(264z_1^4-28z_1^3z_2-86z_1^2z_2^2-21z_1z_2^3-z_2^4)}{(2z_1+z_2)^7(6z_1+z_2)^4}\nonumber \\
&&+\frac{(20z_1^2-30z_1z_2+13z_2^2)}{256z_1(2z_1+z_2)^7}\Biggr\},
\label{ex1s2s}
\end{eqnarray}
and
\begin{eqnarray}
&&{Ex_{1s2p} \over E_0}= -\frac{224}{3}\frac{z_1^3z_2^5}{(2z_1+z_2)^7},
\label{ex1s2p}
\end{eqnarray}
respectively. In addition, the first state is not normalized. Therefore 
\begin{equation}
\langle \Psi_S| \Psi_S \rangle = 1 - 2048z_1^3z_2^3\frac{(z_1-z_2)^2}{(2z_1+z_2)^8}
\label{normS}
\end{equation}
The trial wave function remains normalized for the $^2P$ state, i.e. $\langle \Psi_P| \Psi_P \rangle$ = 1.

If the variational parameters are fixed at $z_1 = z_2 = 3$, the calculation may be considered as a perturbative calculation, where the electron-electron interactions act as the pertubation; these energies are listed in Table \ref{tab:1}. 

\begin{table}
%\begin{ruledtabular}
\begin{tabular}{l c c c c}
 \hline  \hline  \\
& \multicolumn{2} { c }{Perturbative (eV)}&   \multicolumn{2} { c }{Variational (eV)}    \\
\textbf{Term} &  1s$^2$2s  & 1s$^2$2p   & 1s$^2$2s  & 1s$^2$2p  \\
 \hline  \hline  \\
$K$ & 275.5 & 275.5 & 212.1 & 200.0 \\
$U_{nuclear}$ & -551.0 & -551.0 &-486.7 &-459.9 \\
$U_{1s1s}$ & 51.0 & 51.0 & 46.6 & 45.7 \\
$U_{1s2s}$ & 17.1 & - & 11.4 & - \\
$U_{1s2p}$ & - &19.8 & - & 7.1 \\
$Exch_{1s2s}$ & -1.8 & - & 3.9 &- \\
$Exch_{1s2p}$ & - & -1.4  & - & -0.1 \\
$\langle \Psi| \Psi \rangle$ & 1 & 1 & 0.98 & 1 \\
 \hline \\
Total & -192.0 & -186.3 & -201.2 & -200.0 \\
 \hline  \hline  \\
\end{tabular}
%\end{ruledtabular}
\caption{ Energies found for each term using a perturbative calculation ($z_1=z_2=3$) and a 
variational calculation using Eqs. (\ref{1s2s}) and (\ref{1s2p}). In both cases $Z=3$. The 
optimal $z_1$ and $z_2$ values from the variational calculation are given in the text. The
actual values for the $1s^22s$ and $1s^22p$ energies are $-203.5$ eV and $-201.6$ eV 
respectively.\cite{yan95}
}
\label{tab:1}
\end{table}

More accurate energy estimates may be found if $z_1$ and $z_2$ are taken as variational parameters. Minimization of 
Eq. (\ref{1s2s}) yields $z_1 \approx 2.68$ and $z_2 \approx 1.87$, while for the $^2P$ state given by Eq. (\ref{1s2p})
we find $z_1 \approx 2.69$ and $z_2 \approx 1.05$. The proximity of the last number to $Z-2 = 1$, and the 
relative accuracy of the variational calculation (differences of 1.1\% and 0.8\% for the two states) no doubt
combine to inspire the traditional screening argument.

However, these variational calculations implicitly account for electron configurations beyond the $n=1$ and $n=2$ states
under consideration here, More straightforwardly,
inspection of the perturbation-like calculations in Table~1 indicates that the
electron-electron interaction energy between the $1s$ and $2p$ states ($2(U_{1s2p} + Ex_{1s2p}) \approx 36.8$ eV) is
considerably higher than that between the $1s$ and $2s$ states ($2(U_{1s2s} + Ex_{1s2s}) \approx 30.6$ eV).
When one goes beyond perturbation theory, both wave functions expand by incorporating higher energy configurations;
these serve the purpose of allowing the electrons to avoid one another more efficiently, and in so doing, the $1s$ electrons
effectively `screen' the $2s$ or $2p$ electron. So in an attempt to ameliorate
the (rather large) electron-electron energy cost for the $2p$ electron, the variational calculation simply places it
further from the nucleus, and succeeds in lowering this energy, but at a cost of what we have called $U_{\rm nuclear}$
($-551.0$ gets raised to $-459.9$). The same thing happens for the $2s$ electron, but to a lesser extent, in part
because the $2s$ electron is already further extended than the $2p$ electron. Nonetheless, this does not 
succeed in reversing what was evident from the first two columns --- the $1s^22p$ state 
simply has a higher electron-electron energy compared to the $1s^22s$ state, and
therefore the latter has lower energy. 

Phrased this way, it should be evident that `screening' is the `sacrifice' one makes to lower the overall energy. In both the
$2s$ and the $2p$ case, lowering the overall energy was accompanied by screening, i.e. a raising of $U_{\rm nuclear}$ (while
the overall energy decreased). This is qualitatively different than the standard argument that says that the $2p$ state is
more screened to begin with (false!) and therefore its energy is higher.

In summary, we have argued that the ground state electron configuration for Lithium is $1s^22s$, and not $1s^22p$ in
spite of screening, not {\em because of} screening. Expansion of an electron cloud surrounding a nucleus generally occurs
in an effort to minimize the Coulomb repulsion experienced by these electrons.\cite{hirsch02} It follows that
the more repulsive these interactions are, the more will the electrons expand away from one another. In so doing, they also
become more efficiently screened by the remaining electrons; this is a sacrifice that is made in spite of the electron-ion
energy increase, because the total is thereby decreased. The original and incorrect argument stated that the Hydrogenic $2p$
state is more screened than the Hydrogenic $2s$ state, and this is why the latter results in a lower energy. Our argument
clearly attributes the deciding factor to the electron-electron repulsion, and increased screening is a necessary 
(energy raising) by-product that is nonetheless tolerated by the ground state.

\acknowledgments
This work was supported in part by the Natural Sciences and Engineering Research Council of Canada (NSERC), and by the Teaching and Learning Enhancement Fund (TLEF) at the University of Alberta.


\begin{thebibliography}{0}
\bibitem{remark0} We single out the ion(s) in this discussion because they are much more massive 
than the electrons; in other words we assume a Born-Oppenheimer approximation.
\bibitem{remark2} Note that in the case of Helium, while the second electron is primarily in the $1s$ state
because that minimizes the electron-ion energy, there is nonetheless almost 10\% probability that either
electron is {\em not} in the $1s$ level, specifically because this lowers electron-electron interaction energy.
Thus in the case of Helium, screening (i.e. some of the electron cloud is pushed further out) occurs to lower the energy, not the
other way around.
\bibitem{griffiths289} See Ref. [\cite{griffiths05}], p. 289, problems 6.34 and 6.35, where these
relations are derived.
\bibitem{remark3} The existence of the wall at $r=0$ does not effect the solution since the radial
wave function $u(r) \equiv rR(r)$ is required to be zero there anyways.
\end{thebibliography}
\end{document}